\documentclass[aps,prd,twocolumn,showpacs,superscriptaddress]{revtex4}

\usepackage{epsfig}
\usepackage{graphicx}
\usepackage{enumitem}

\begin{document}

\title{On the possibility of Vacuum-QED measurements with gravitational wave detectors}

\author{H. Grote}
\email{hartmut.grote@aei.mpg.de}
\affiliation{Max-Planck-Institut f\"ur Gravitationsphysik (Albert Einstein Institut) und
Leibniz Universit\"at Hannover, Callinstr.\,38, 30167 Hannover, Germany}

\raggedbottom

\date{\today}

\begin{abstract}
Quantum electro dynamics (QED) comprises virtual particle production
and thus gives rise to a refractive index of the vacuum
larger than unity in the presence of a magnetic field.
This predicted effect has not been measured to date, even
after considerable effort of a number of experiments.
It has been proposed by other authors to possibly use 
gravitational wave detectors for such vacuum QED measurements, 
and we give this proposal some new consideration
in this paper. In particular we look at possible source field 
magnet designs and further constraints on the implementation at
a gravitational wave detector. 
We conclude that such an experiment seems to be feasible
with permanent magnets, yet still challenging in its implementation.

\end{abstract}


\maketitle

\section{Introduction}
\label{introduction}

Corrections to the Maxwell equations that emerge from the quantum properties
of the vacuum have been proposed many decades ago, see e.g. \cite{Heisenberg1936}.
Quantum electrodynamics (QED) predicts that the velocity of light
propagating in vacuum is decreased in the presence of a magnetic field.
In particular, a light ray traversing a region with a magnetic 
field $B$ with its field lines oriented perpendicular to the light propagation
direction and parallel to the polarization direction of the light,
should slow down due to an increase of the refractive index of 
\begin{equation}
\Delta n_{\parallel}=9.3\times10^{-24} \times B^2 [1/T^2] ,
\end{equation}
as for example derived in \cite{Doebrich2009} and some references therein.
Here $B$ denotes the magnetic field strength (in units of Tesla) 
traversed by the light. If the magnetic field is oriented perpendicular
to the polarization direction of the light, a smaller increase of the
refractive index of 
\begin{equation}
\Delta n_{\perp}=5.3\times10^{-24} \times B^2 [1/T^2]
\end{equation}
is predicted~\cite{Doebrich2009}.
Given these results, we can define the difference between $\Delta n_{\parallel}$ and
$\Delta n_{\perp}$ as
\begin{equation}
\Delta n_{\parallel-\perp} = \Delta n_{\parallel} - \Delta n_{\perp} =
4\times10^{-24} \times B^2 [1/T^2]
\label{eqdeltanpp}
\end{equation}

To date, this fundamental prediction of QED is still unconfirmed, 
even though a number of experiments have tried or are trying 
to measure the effect, most notably PVLAS and 
BMV~\cite{Valle2010, Valle2013, Valle2014b, Berceau2012, Cadene2014}, 
but also Q\&A~\cite{Chen2007}. 
Others have been proposed, as e.g. OSQAR~\cite{Pugnat}
and a pulsed laser experiment~\cite{Heinzl2006}.

All of the ongoing experiments make use of the difference $\Delta n_{\parallel-\perp}$ 
of the predicted refractive index changes for different angles of the magnetic field
with respect to the polarization direction of the light,
i.e. they attempt to measure the \emph{birefringence} of the vacuum.
In these experiments, a laser beam resonating in a Fabry-Perot cavity
passes a magnetic field, resulting in different refractive
indices for the two orthogonal polarization directions.
Ellipsometers then measure a rotation of the polarization
of the light as a measure of the vacuum birefringence. 
In PVLAS and Q\&A a modulation of the \emph{angle} of the magnetic field with
respect to the polarization direction of the light is used (and thus a modulation of the
induced polarization rotation) to suppress effects at low frequencies 
and isolate the measured signal from background noise.
Not yet understood excess noise from birefringence of highly-refractive
mirrors has led to problems in the past, resulting in a `signal' above 
the expected QED signal e.g. in the PVLAS experiment. While this could be explained 
later, the birefringence of highly reflective mirrors 
remains a problem to date, see e.g. \cite{Bielsa2009} and references therein.
The experimental upper limit for birefringence of the vacuum 
established by BMV~\cite{Cadene2014} is still a factor of about 2000 away 
from the predicted value.
The new PVLAS experiment could recently significantly improve the
upper limit to a factor of 50 above the prediction~\cite{Valle2014b}.

While the PVLAS experiment uses a static magnet (e.g. a permanent magnet
in the new design~\cite{Valle2013}), 
another approach gained momentum with the notion that higher magnetic
fields, and thus larger signals, could be produced with pulsed magnets.
Pulsed magnets modulate the \emph{amplitude} of the magnetic field,
and thus - by nature of the pulses - provide a modulation to suppress 
background noise as well.
Askenazy et.al. were the first to propose a pulsed coil design for
measurements of birefringence~\cite{Askenazy2001}.
The BMV experiment~\cite{Battesti2008} started using a
pulsed coild design slightly different from this, called \emph{xcoil} 
~\cite{Batut2008}, in order to approach the measurement of vacuum 
birefringence with pulsed magnets. As we will see below, the larger 
signal from pulsed magnets has to be balanced against integration
time.

A different approach to the measurement of vacuum-QED effects
was mentioned in~\cite{Boer2002}, namely
to use laser interferometers for gravitational wave (GW) detection, to measure directly
the \emph{velocity shift} (rather than the \emph{polarization shift}) 
of the light in the presence of a dedicated magnetic field.
This may be the first time that GW detectors have been mentioned explicitly
in this context, however the idea to measure the velocity shift of light in the
presence of a magnetic field has been proposed several times earlier, as pointed out
in an excellent overview article by Battesti and Rizzo~\cite{Battesti2013}.
A nice example of this is the paper by Grassi Strini, Strini, and Tagliaferri from 1979, who
already discussed the use of laser interferometers for vacuum-QED 
measurements~\cite{Strini1979}.
The proposal to use GW detectors for QED measurements was also picked up in \cite{Denisov2004}, 
where the authors come to the conclusion that prototype GW-interferometers would be more suitable
than full-scale GW-detectors. However, this conclusion is incorrect due to
a false assumption on how the interferometer displacement noise scales with
an increase of arm-cavity Finesse. 

Later, Zavattini and Calloni, pointing out the error in \cite{Denisov2004}, 
studied some implications of attempting vacuum-QED measurements
for the case of the Virgo interferometer ~\cite{Zavattini2009}. 
They consider the use of dipole magnets to be used quasi-continuously
at a fixed frequency, and put forward some more principal considerations of how
such a magnet could be incorporated into the GW-detector.

D\"obrich and Gies ~\cite{Doebrich2009} then proposed to use
\emph{pulsed magnets} to measure the velocity shift of the light
in gravitational wave detectors. 
They point out that not only do pulsed magnets yield larger signals for the same amount 
of average energy driving the magnet, but also naturally can match the frequency 
response of gravitational wave detectors in a potentially favorable manner.
I will get back to these considerations in section \ref{subsec_pulsed}.

Table \ref{overviewtable} shows an overview of existing or considered
vacuum-QED measurements, arranged by the modulation method of the magnetic
field and the measured quantity of the affected light.
It was pointed out in particular by Zavattini and Calloni in~\cite{Zavattini2009},
that the independent measurement of $\Delta n_{\parallel}$ and $\Delta
n_{\perp}$ allows to distinguish between different possible particle
models, in case a signal larger
than the expected vacuum-QED effect would be observed.

\begin{table}[h]
\begin{center}
\begin{tabular}{|l||c|c|}
\hline
 & \parbox[t]{2.5cm}{Rotate B-field\\} & \parbox[t]{2.5cm}{Modulate B-field amplitude\\}\\
\hline
\hline
\parbox[t]{2.5cm}{Measure polarization\\} & PVLAS, Q\&A, others & BMV \\
\hline
\parbox[t]{2.5cm}{Measure velocity\\} & GW detectors & \parbox[t]{2.5cm}{GW detectors,  more physics\\}\\
\hline
\end{tabular}
\caption{Overview of existing or considered vacuum-QED measurement attempts, arranged
by the manipulation method of the magnetic field and the measured quantity. 
Only in the case of amplitude-modulating the magnetic field \emph{and} measuring the 
resulting velocity modulation of light, 
can $\Delta n_{\parallel}$ and $\Delta n_{\perp}$ be measured independently,
which is denoted as \emph{more physics}.}
\label{overviewtable}
\end{center}
\end{table}

To date none of the proposals to use GW detectors for vacuum-QED
measurements covers in detail the discussion
of a possible magnet design.
Trying to fill this gap is the main aim of this paper.
In section \ref{general}, we look at some general considerations on the
requirements of a magnetic field for vacuum-QED measurement purposes.
We give an expression for the signal integration time as a function of the 
signal amplitude over detector noise, and discuss this in the light of expected noise levels of 
(near) future GW-detectors.
In section \ref{magnets} we discuss different possible magnet types for
a given example scenario.  
The use of permanent magnets is identified as the most favorable source of 
the magnetic field, and in 
section \ref{realistic} we look in more detail at a possible realistic setup
using permanent magnets.

\section{General considerations}
\label{general}

Laser-interferometric gravitational wave detectors are
ultra-sensitive length measurement devices which push several technologies
to their limits in order to reach displacement sensitivities of order
$10^{-20}\,\rm m / \sqrt{Hz}$ around 100\,Hz and 
below~\cite{Harry2010,AdvVirgo2009,Kagra2012,GEO-HF}. 
In these instruments, a Michelson interferometer configuration 
(with some optical enhancements) is used to measure
differential length fluctuations between two perpendicular laser
beam paths. A passing gravitational wave causes differential length 
perturbations between the two paths, 
which result in a phase shift of the light beams that
can be detected upon re-combination at the beam splitter of the
Michelson interferometer.
This basic functionality may open up the possibility to use
the exquisite sensitivity to length changes 
(or equivalently: sensitivity to phase shifts of light)
for fundamental physics measurements in addition to the
primary purpose of the detection of gravitational waves.

To consider the feasibility of vacuum-QED measurements using GW detectors,
we need to relate possible (QED) signal sizes to the sensitivity of GW detectors.
We calculate the signal $S_{\parallel}$ we obtain from a magnetic-field induced change of
the refractive index as
\begin{equation}
S_{\parallel} = \Delta n_{\parallel} \times D = 9.3\times10^{-24} \times B^2 [\frac{1}{T^2}] \times D\,,
\label{signal}
\end{equation}
with D being the effective length over which the magnetic field B
is applied. As will be seen below, we have to apply a modulation in time
to the field B, of the form $B(t)=B_0 cos(\omega t)$, in order to be able to measure 
it with a GW detector.
We then obtain:
\begin{equation}
S_{\parallel} = \Delta n_{\parallel} \times D = 9.3\times10^{-24} \times B_0^2 \times \frac{1}{2} (1+cos(2 \omega t)) [\frac{1}{T^2}] \times D,
\label{signal2}
\end{equation}
with a signal $S$ at twice the modulation frequency $\omega$.
We note that $S$ is a sinusoidal signal for which a convention must be
used how to denote its amplitude (in case the time dependence is omitted). 
While $B_0$ in equation~\ref{signal2} denotes the peak amplitude of the
exciting field, we obtain peak-to-peak values for $S_{\parallel}$ due to the
squaring of $B(t)=B_0 cos(\omega t)$. 

Since the signal $S$ has the units of meters, it is most
natural to convert the gravitational-wave strain sensitivities
typically given for GW detectors to displacement 
sensitivities~\footnote{It is not necessary to incorporate the discussion 
of other technical parameters of the interferometer, as for example the
arm-cavity finesse, as has been done 
by other publications, leading to erroreonous conclusions in few cases.
It is worth noting that the calibration of GW detectors is actually done
in the way to apply a known displacement to one or more test-masses,
and thus calibrate the output directly in meters. Only after this step, the strain
sensitivity to gravitational waves is obtained by dividing the displacement
by the interferometer arm length.}.

The definition of GW-strain $h$ is $h=2\frac{\Delta L}{L}$, where
$\Delta L$ is displacement (or GW-induced length change) applied 
to \emph{each} arm in a differential manner, and $L$ is the length
of each interferometer arm (assuming equal length for both arms). 
The differential nature of the length change of the two interferometer 
arms is inherent to a gravitational wave.
However, if we want to use the GW detector to measure length changes
of a \emph{single} arm (i.e. by applying a magnetic field to just one arm), 
we simply obtain $h_{eq}=\frac{\Delta L_x}{L}$.
Here $h_{eq}$ denotes the equivalent quantity to compare to GW strain $h$
(to which the detector is calibrated), and $\Delta L_x$ denotes the length
change of a single 
arm~\footnote{The same argument is derived in \cite{Zavattini2009}.}.
With this we obtain $\Delta L_x = L h_{eq}$, where we can interpret
$\Delta L_x$ as the length change of a single interferometer arm of length $L$,
given a GW-strain equivalent of $h_{eq}$. 
We therefore can multiply a GW-detector strain spectrum with the arm
length of the detector, and get a displacement spectrum which we can compare 
to a displacement signal generated in one arm.

If we want to compare the signal $S$ to noise
curves of GW-detectors, we take note of the fact that the GW-detector
noises are given as $RMS$ values of a sinusoid.
To translate $S_{\parallel}$ into a signal $S_{RMS,\parallel}$,
we need to apply another factor:
\begin{equation}
S_{RMS,\parallel} = \frac{S_{\parallel}}{2 \sqrt 2 } 
\label{pp-rms}
\end{equation}
The factor $2 \sqrt 2$ comes
from the fact that equation \ref{signal2} yields a \emph{peak-to-peak} value
of the signal for the modulated sinusoidal field $B(t)$.

Assuming continuous application of a sinusoidal signal $S_{RMS,\parallel}$ 
we calculate the integration time needed to obtain a signal-to-noise 
ratio (SNR) of $SNR=1$ as
\begin{equation}
t_{SNR=1} = \left(\frac{\tilde n(f)}{S_{RMS,\parallel}}\right)^2,
\label{inttime}
\end{equation} 
where $\tilde n(f)$ is the displacement noise amplitude spectral density
of the length measurement device (GW detector) at a frequency $f$
of choice.
 
To distinguish the signal $S$ from the product $B^2 \times D$ which
relates to the magnet strength and interaction length,
we define the \emph{excitation} $E$ as
\begin{equation}
E := B^2 \times D
\label{excitation}
\end{equation} 
This will be the main quantity to maximize for a given magnet setup.

\begin{figure}[htbp]
\centering
\includegraphics[width=1.0\linewidth]{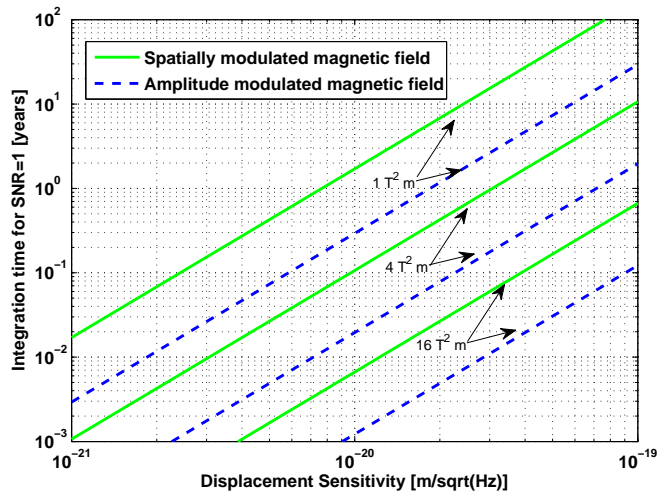}
\caption{Integration times for a signal-to-noise ratio of unity,
as a function of GW detector sensitivity (given as \emph{rms} values). 
The lines denote different excitation strengths (as \emph{peak-to-peak} values) 
for spatially- and amplitude modulated fields, respectively.
Continuous application of a sinusoidal signal is assumed.
}
\label{inttimes}
\end{figure}
\begin{figure}[htbp]
\centering
\includegraphics[width=1.0\linewidth]{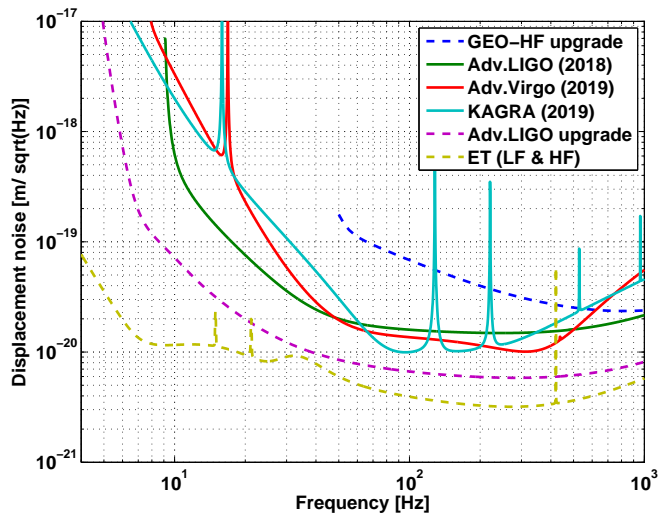}
\caption{Planned displacement noise curves for laser-interferometric
gravitational-wave detectors. Advanced LIGO~\cite{Harry2010},
Advanced Virgo~\cite{AdvVirgo2009}, and KAGRA~\cite{Kagra2012} are 
under construction, and are denoted with solid lines.
The dashed lines denote potential upgrades to GEO-HF~\cite{GEO-HF} 
and Advanced LIGO,
as well as the proposed Einstein telescope ET~\cite{ET2011}. 
Years denote
estimated/hypothetical times of reaching the target sensitivity.
As customary in this field, the noise curves are displayed as 
root-mean-square amplitude spectral densities.}
\label{alldetectors}
\end{figure}

Figure~\ref{inttimes} shows integration times for a SNR of unity,
as a function of GW detector sensitivity, according
to eq.~\ref{inttime} with continuous application of a sinusoidal 
signal $S$ being assumed.
The graph in Figure~\ref{inttimes} shows two lines each, for three different 
excitation strengths $E$. 
The solid line assumes the rotation
of a static magnetic field around the laser beam axis and thus is suitable
to measure $\Delta n_{\parallel - \perp}$. 
The dashed line assumes an amplitude modulation of the magnetic field,
and thus is suitable to measure $\Delta n_{\parallel}$, for a parallel 
orientation of the magnetic field and the polarization of the light field. 

The integration times calculated from eq.~\ref{signal2}-\ref{inttime}
and shown in Figure~\ref{inttimes}
are a factor of two longer than the times calculated
in reference~\cite{Zavattini2009} for identical setups. This discrepancy comes from 
an incorrect assumption on the calibration of GW interferometer strain 
data~\cite{Zavattini2013}.

Finally, Figure~\ref{alldetectors} shows planned displacement sensitivity curves
for ground-based laser-interferometric gravitational-wave detectors 
and potential upgrades.
The projects Advanced LIGO~\cite{Harry2010}, Advanced Virgo~\cite{AdvVirgo2009}, and
KAGRA~\cite{Kagra2012} are currently under construction and are
expected to become operational within the next years.
Other projects are potential upgrades~\cite{Sred2012} or entirely new detectors,
as in case of the Einstein Telescope ET~\cite{ET2011}.

To give an example, for Advanced LIGO and Advanced Virgo
we can read a sensitivity of $2\times10^{-20}\,\rm m/\sqrt{Hz}$ at 50\,Hz.
With an excitation of $1\,\rm T^2 m$ for an amplitude-modulated magnetic field
(which would have to be modulated at 25\,Hz), 
we get an integration time of a bit more than 1 year for a SNR of unity,
according to Figure~\ref{inttimes}.

While this integration time seems very long, it should be noted that
long integration times pose no principal problem here, since
the gravitational-wave detectors are expected to run for several 
years. Obviously, a magnetic field excitation would have to be held active
during this time as well.
However, such a long integration time would probably be the upper acceptable
limit for SNR=1, and either more sensitivity or a stronger field excitation would be
desirable in the long run.
In the following,
we will use the example of a magnet system with $1\,\rm T^2 m$ for amplitude modulated fields,
and the example of $2.3\,\rm T^2 m$ for rotating fields, which gives
similar integration times for the measurement of $\Delta n_{\parallel}$
and $\Delta n_{\parallel - \perp}$, respectively.

\section{Discussion of possible source field magnets}
\label{magnets}

To illustrate the difficulty of building large strong magnets,
it is instructive to calculate
the energy stored in a magnetic field, $W = \frac{1}{2 \mu _0} B^2 V$,
with $\mu _0$ being the permeability of the vacuum, and $V$ the volume over
which the magnetic field $B$ is erected. If we assume the volume will be of
cylindrical shape with radius $r$
around the laser beam axis and extending over a length $D$,
we get for the energy $W$:

\begin{equation}
W = \frac{\pi}{2 \mu _0} B^2 D r^2
\label{eq_energy}
\end{equation}

We see that the energy content in the magnetic field increases with the square of the
field strength \emph{and} with the square of the field radius, which is one reason
why the design of large \emph{and} strong magnets is technically challenging.

In order to discuss possible source field magnets which are as small as
possible, but as large as necessary for our application, we need to 
determine the minimum radius $r$ of the usable magnetic aperture 
through which the laser beam would pass. 
GW detectors have the laser beams traversing
in stainless-steel beam tubes of order 1\,m in diameter.
However, as has been discussed in~\cite{Zavattini2009}, possibly a smaller
aperture has to be used for vacuum-QED measurements.
We propose here to use the smallest aperture possible, as judged by the constraints
of the GW detectors' optical path. Therefore, we note the 
additional loss that an aperture 
(due to a section of the beam tube with reduced diameter) would cause.
GW detectors use laser beams with a Gaussian beam profile, having a radial intensity
distribution of 
\begin{equation}
I_{(r)} = I_0 e^{\frac{-2r^2}{w^2}}.
\end{equation}
Here $I_0$ is the intensity at the center of the beam ($r=0$) and $w$ is the
(1/e field-) beam radius. 
From this, the power loss due to clipping of the beam profile is
calculated.
However, an aperture does not only clip the beam, it also gives rise
to diffraction. The laser power diffracted from the central gauss-beam profile
is not recovered by the optical resonators within the GW detector and thus lost.
The effective loss from diffraction slightly depends on the geometry of
the optical resonator within which the aperture is located, such that a
simulation~\cite{FINESSE} has been used here, for the example of Advanced LIGO.
The results of this simulation correspond to those obtained in~\cite{Spero}
when adjusting for the wave-length and cavity geometry used in there.

\begin{figure}[htbp]
\centering
\includegraphics[width=0.7\linewidth]{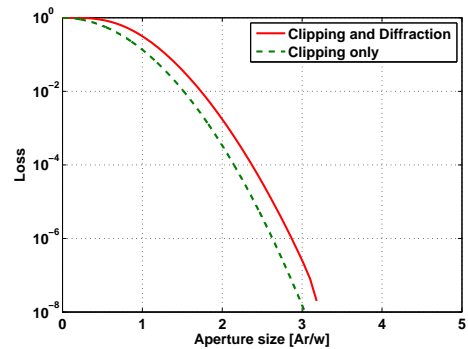}
\caption{Calculated power loss of a laser beam with Gaussian beam profile 
with radius $w$, as a function of the size of a circular beam clipping
aperture with radius $A_r$ (located at the beam waist). 
Losses due to clipping and total loss (i.e. clipping and diffraction)
are shown separately.}
\label{beamclipping}
\end{figure}

Figure~\ref{beamclipping} shows the calculated power loss of a 
laser beam with Gaussian beam profile as a function of the size of 
a circular beam clipping aperture. The pure clipping loss is shown separately
from the total loss obtained from the simulation.
Advanced GW detectors are designed to have very low optical losses upon
reflection of laser beams on their mirrors. These low losses which
should be no more than about 30-50\,ppm per reflection 
are mandatory to allow
a high power build-up in the resonant optical cavities.
Therefore, and also to minimise scattered light from the aperture, no more than
around $1\,ppm$ of extra loss from a reduced tube aperture
seems acceptable.
Given Figure~\ref{beamclipping}, an aperture radius of 3 beam radii appears as 
a reasonable choice, yielding about 0.2\,ppm (parts per million) loss.


Table~\ref{aperturetable} shows actual beam sizes of current and planned
GW detectors, around the middle of the beam tube, near to the position of
the minimum beam size. (GEO\,600 is an exception to this, since in the current
layout the minimum beam size is located close to the corner-station end of the beam tube.)
\begin{table}[h]
\begin{center}
\begin{tabular}{|l|c|c|c|}
\hline
GW-IFO & \parbox[t]{2cm}{Beam radius at waist [mm]\\ } & 
\parbox[t]{2cm}{Minimum aperture\\(3 x beam radius) [mm]\\ } &
\parbox[t]{2cm}{Realistic aperture\\ radius [mm]\\ } \\
\hline
\hline
GEO\,600 & 9 & 27 & 40 \\
\hline
Adv. Virgo & 10 & 30 & 45 \\
\hline
Adv. LIGO & 12 & 36 & 55 \\
\hline
KAGRA & 16 & 48 & 70 \\
\hline
ET-HF & 25 & 75 & 115 \\
\hline
ET-LF & 29 & 87 & 130 \\
\hline
\end{tabular}
\caption{Beam radii of existing and planned gravitational-wave detectors, and
proposed minimum aperture sizes. In the column 'realistic aperture radius', 
50\,\% is added to the minimum aperture and the result rounded. 
This accounts for an additional beam tube and
clearance space, as detailed in section~\ref{realistic},
however is only an example here, since exact numbers would depend on more details
of a chosen setup.}
\label{aperturetable}
\end{center}
\end{table}

As an example, aiming for an aperture radius of 55\,mm and an excitation of $E
= 1\,\rm T^2 m$, we calculate the 
energy stored in the magnetic field of such a magnet to be
$W = 3781\,\rm J$.
For continuously amplitude-modulated fields with a modulation frequency of 25\,Hz,
this energy has to be brought into and removed from the aperture space
50 times per second, corresponding to an energy flow of almost 200\,kJ/sec.

\subsection{Electro-magnets}
\label{subsec_electro}

\begin{figure}[htbp]
\centering
\includegraphics[width=1.0\linewidth]{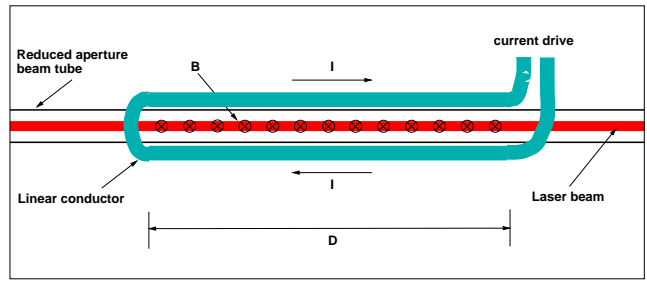}
\caption{Principal setup of a linear magnet consisting of a conductor
loop running mostly in parallel to both sides of a laser beam tube.
Some quantities are denoted as used in the text.}
\label{Linear1}
\end{figure}

Electro-magnets, modulated in field amplitude, would allow for the measurement of
$\Delta n_{\parallel}$ or $\Delta n_{\perp}$. 
Linear conductors parallel to the laser beam tube are
the most efficient way to generate a magnetic field perpendicular to the laser
beam direction, a setup also used for beam deflection in particle accelerator
magnets. 
Figure~\ref{Linear1} shows a principal setup of a linear magnet arranged
alongside a laser beam tube.

We start with the parameter estimation of such a linear
conductor, by noting that we want to maximize the excitation $E$,
as defined in equation~\ref{excitation} for any magnet design.
Eq.~\ref{excitation} is to be resolved to the geometric parameters of the setup,
material constants, and electrical power.
We use the following equations:
\begin{equation}
\label{eqa3}
B = 2 \mu _0 \frac{I}{2 \pi r}
\end{equation}
with $I$ being the effective current through the linear conductor
with distance $r$ from the laser beam axis. This is an approximation
assuming a conductor which is long in the laser beam direction, 
and has a small crossection compared to the
dimensions of the beam tube in the plane perpendicular to the beam tube.
The factor of two on the left side comes from counting the two conductors
at each side of the beam tube.
We then use the electrical power dissipation $P$ (in the low frequency limit)
\begin{equation}
\label{eqa4}
P = I^2 R
\end{equation}
with $R$ being the total ohmic resistance of the linear conductor.
Finally we use
\begin{equation}
\label{eqa5}
R = \frac{2 \rho D}{A}
\end{equation}
to calculate the ohmic resistance of the conductor, with $\rho$ being the 
specific resistance of the conductor material and $A$ being
the cross-section of the conductor~\footnote{
A possible splitting of the cross-section A into divisions of several coil
windings of a conductor with smaller cross-section
is not taken into account here, but does in fact not change this basic result.
The number of windings is only important to adjust to further technical
parameters, as e.g. the ratio of current to voltage.}.

Combining equations (\ref{eqa3}) to (\ref{eqa5}) and inserting into (\ref{excitation}) 
we obtain
\begin{equation}
\label{eqa6}
E = \frac{\mu _0 ^2}{2 \pi ^2} \frac{P A}{\rho r^2}
\end{equation}
This result has the following implications for the magnet design:

\begin{itemize}

\item{The length of the conductor $D$ has been eliminated and
hence has no influence on the excitation strength (for a given power,
conductor cross-section, and beam tube diameter).
Note that this result can be used to adjust the maximal temperature as well as
mechanical force on the conductors, as two technical constraints.}

\item{For a fixed tube diameter, the excitation increases with increasing 
conductor cross-section, up to practical limits not reflected in eq.~\ref{eqa6}. 
In principle this can be used for the estimation of a quasi-optimal
conductor cross-section.}

\item{Increasing the electrical power $P$ and decreasing the resistance
of the conductor material $\rho$ increase the excitation linearly.
While copper is the obvious material of choice for the conductor, the power $P$ will be
determined by heat dissipation and general power handling constraints of the setup.}

\item{Similar to what was found in eq.~\ref{eq_energy}, the excitation
decreases with the square of the distance $r$ to the application region,
which thus should be as small as possible.}

\end{itemize}

For a desired excitation of $E = 1\,\rm T^2 m$ and an example conductor 
cross-section $A \approx r^2$ we obtain (using eq.~\ref{eqa6}) a necessary
power of $P \approx 210\,\rm kW$. 
This is a rather large power and seems very impractical to realize.

\begin{figure}[htbp]
\centering
\includegraphics[width=0.6\linewidth]{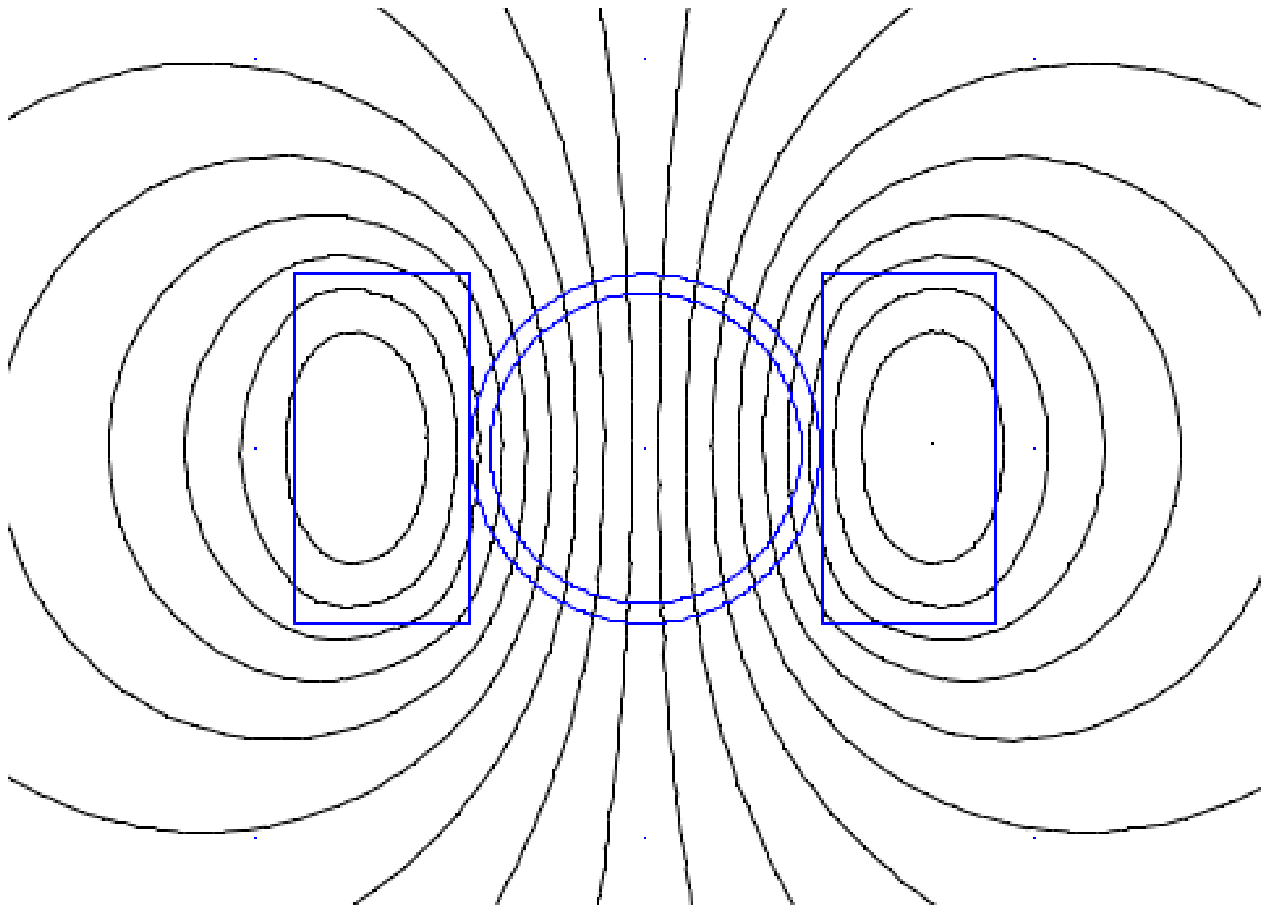}
\caption{Setup and simulated magnetic field lines for a linear conductor of rectangular
  shape to both sides of the central circular vacuum tube. 
This is the cross-sectional view of the setup as in Figure~\ref{Linear1}.}
\label{Mag1}
\centering
\includegraphics[width=0.6\linewidth]{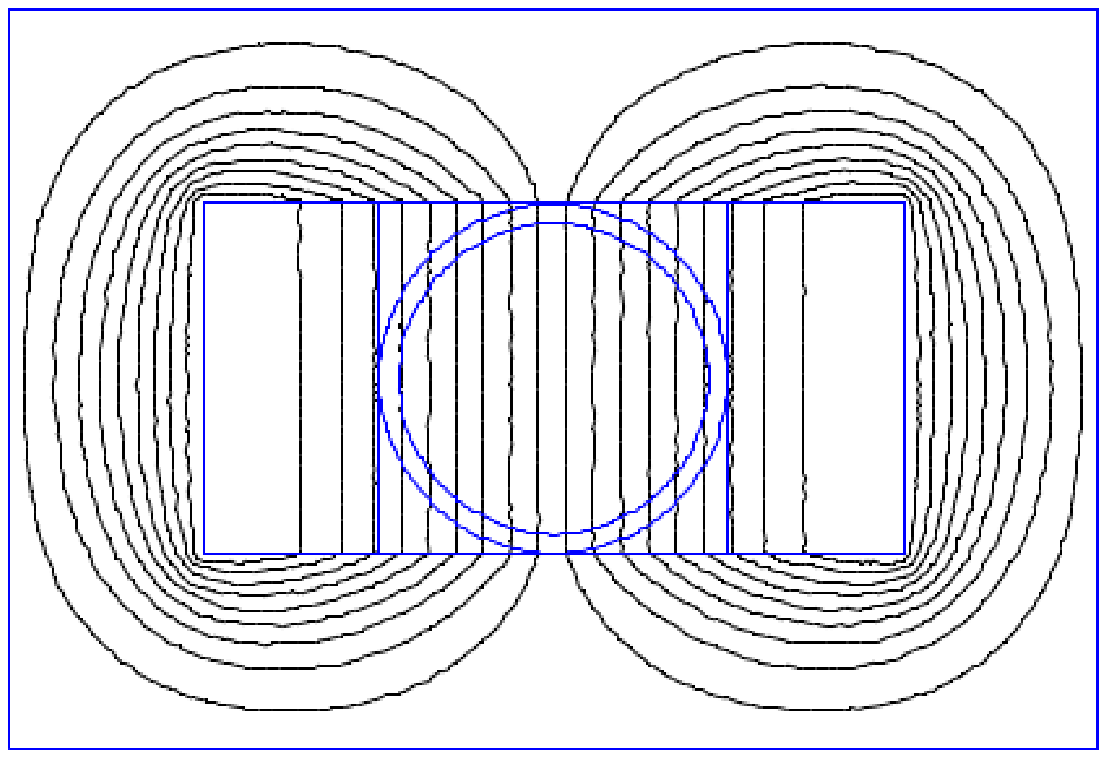}
\caption{Setup and simulated magnetic field lines for the same setup as in
  Figure~\ref{Mag1}, but with the addition of ferro-magnetic material
around the rectangular conductors and the central vacuum tube.}
\label{Mag2}
\end{figure}

A simulation with the finite-element simulator program \emph{FEMM}~\cite{FEMM}
yields $P \approx 300\,\rm kW$, roughly confirming the simplified calculation.
Just for illustration, Figure~\ref{Mag1} shows the magnetic field
lines for this simulation.
However, this result is only valid for the low frequency limit,
in particular at DC. 

While super-conducting magnets can lower the energy dissipation
in the conductor due to the extremely low electrical resistance, 
they are not suitable for large and fast amplitude modulations 
of the magnetic field, as will be required for our application. 
Therefore, we only consider copper as conductor material throughout this paper.

For our example, we need to amplitude-modulate the drive current 
at a frequency $f=25\,\rm Hz$.
While the \emph{real} power dissipation is largely un-affected from this 
(neglecting skin- and proximity effects), the inductance of the setup
results in very large \emph{reactive} powers to be handled.
The numerical calculation with FEMM yields a reactive power of $P=2.5\,\rm MW$,
which very much complicates the electric drive circuit on top of the real
power dissipation of the system as calculated above.

\subsubsection*{Field enhancement with ferro-magnetic material}

If the peak magnetic fields are constrained to about 2\,T, it can be considered
to use a ferro-magnetic material to enhance the magnetic flux for a given
magnetic setup. 
This has been simulated for the example above,
adding a soft ferro-magnetic material (US-steel type S-2, with 0.018\,inch
lamination) in the FEMM simulation around the copper conductor,
as depicted in Figure~\ref{Mag2}.
The electrical drive power has been adjusted to again yield an excitation
of $E = 1\,\rm T^2 m$. The simulation result is that in the low frequency limit 
the power is reduced to
$P=45\,\rm kW$. 
If modulated at a frequency of $f=25\,\rm Hz$ the reactive power is
now $P=1\,\rm MW$. This is better, but still seems far from practical to realize.

\subsection{Pulsed electro-magnets}
\label{subsec_pulsed}

As will be shown in the following, 
continuous operation of an electro-magnet (at a fixed frequency) is not
optimal. For the same average power $P$ applied to the magnet, the
total integration time for a given SNR can be reduced if the magnet
is active only for fractions of time, intersperced by pauses.

If equation~\ref{inttime} is combined with equation~\ref{eqa6}, we obtain
\begin{equation}
\label{eqa7}
t_{SNR=1} \sim \left(\frac{\tilde n(f)}{P}\right)^2
\end{equation}

We define $P = P_p \times \eta _p$ with $P_p$ being the power applied
to the magnet during a fraction $\eta _p$ of the time, such that 
the average power $P$ is kept constant.
After a few steps we then get
\begin{equation}
\label{eqa8}
t_{SNR=1} \sim \eta _p \times \left(\frac{\tilde n(f)}{P}\right)^2
\end{equation}

Notably, this is the same result as in equation~\ref{eqa7}, except for the factor 
$\eta _p$, which implicitly is $\eta _p = 1$ for the case of continuous
operation of the magnet.
We see that the integration time is linearly reduced with the fraction
of time that the magnet is engaged (keeping the average power $P$ constant).
This is the basic motivation to use a pulsed operation of electro-magnets
in the first place.
Ultimately this technique is limited by the peak power and pulse energy 
that can be handled by the system, which is determined by the pressure 
on the conductors due to Lorentz forces and by constraints in the electrical 
drive system.
Another limit on $\eta _p$ comes from the usable signal period $T=1/f$, 
which preferably has to match the frequency of lowest noise of the 
GW detector, as relevant for our application.
However besides these more technical constraints, 
we also have neglected so far the energy $W_m$ needed to build up the
magnetic field.
To include this energy into the calculation we split the total energy $W_p$ of 
a single pulse into the components due to electrical dissipation in the conductor
$W_E$ and the energy in the magnetic field $W_m$. For $W_P=W_E+W_M$ we obtain:
\begin{equation}
\label{eqa9}
W_P = \frac{L}{2}I^2 + t_P R I^2 
\end{equation}
with $L$ being the inductance of the conductor setup, and $t_P$ being the length of
the pulse. For a linear conductor setup according to Figure~\ref{Linear1} the
inductance L can be approximated as $L=D \mu _0 / \pi$. Resolving equation
\ref{eqa9} to the current $I$, inserting into equations ~\ref{eqa3} and then
into ~\ref{excitation}, we obtain (with also using equation~\ref{eqa5}):
\begin{equation}
\label{eqa10}
E = \frac{\mu _0 ^2}{\pi ^2 r^2} \frac{W_P}{\frac{\mu _0}{2 \pi} + \frac{2 \rho}{A}t_P}
\end{equation}

This result is equal to equation~\ref{eqa6}, if the constant term 
$\frac{\mu _0}{2 \pi}$, which resembles the energy stored in the magnetic field,
is neglected. The implications are:
\begin{itemize}

\item{If the cross-section $A$
of the conductor is small, the energy dissipation $W_E$ will dominate over
the energy $W_M$ in the magnetic field. This is a typical operation regime
for most pulsed magnets.}

\item{If the pulse duration $t_P$ gets too short, $W_M$ will dominate
over $W_E$ such that the excitation $E$ is not increasing any more 
for shorter pulses.
}

\item{Obviously the excitation $E$ scales with the total pulse energy
$W_P$, and inversely with the square of the system size $r$.
}

\end{itemize}

A single pulse of length $t_P$ with energy $W_P$ results in a power $P_P=W_P/t_P$.
To keep the average power $P$ at a fixed lower level we apply pulses only every
$t_p/\eta _p$ seconds, again with $\eta _p=P/P_P$. In order to calculate integration times according to
equation~\ref{inttime}, we then have to use
\begin{equation}
t_{SNR=1, pulsed} = \frac{t_{SNR=1}}{\eta _p}
\label{inttime2}
\end{equation} 

\begin{figure}[htbp]
\centering
\includegraphics[width=1.0\linewidth]{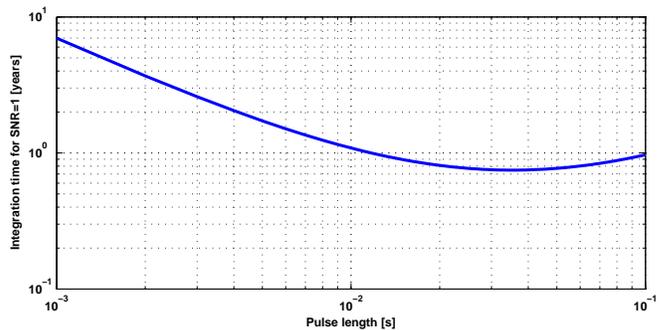}
\caption{Integration times as function of pulse length. The parameters used
  are a pulse energy of $W_P=1\,\rm MJ$, aperture radius $r=55\,\rm mm$, and a
conductor cross-section of $A \approx r^2$.
A GW detector sensitivity of $\tilde n(f)=2\times 10^{-20}\, \rm m/\sqrt{Hz}$
is used for all frequencies}
\label{Pulsed1}
\end{figure}

For a pulse energy of 1\,MJ, an aperture radius of $r=55\,\rm mm$, a
conductor cross-section of $A \approx r^2$, and an average power of 
$P=20\,\rm kW$ we calculate integration times for
SNR=1 as shown in Figure~\ref{Pulsed1}.
A GW detector sensitivity of $\tilde n(f)=2\times 10^{-20}\, \rm m/\sqrt{Hz}$
is used for all frequencies, in order to illustrate the effect of the energy
in the magnetic field on the optimal pulse length~\footnote{When optimizing
the pulse length for a particular detector, the frequency dependence of the
detector sensitivity, according to Figure~\ref{alldetectors}, has to be taken into account.}.

For a pulse length of $t_p=10\,\rm ms$ we get an integration time around 1 year,
which means that about 600000 pulses would have to be applied.
While the average power has been lowered compared to the examples above,
a pulse energy of 1\,MJ is on the edge of current technology and is
a very optimistic assumption given that the magnet for this application
would still need to be developed. An energy of 1\,MJ corresponds to 240\,g
of the explosive TNT.

Further, for the estimations in this section we have made simplifying assumptions,
particularly not taking into account the following parameters:
forces between the conductors and within the conductors,
temperature rise of the conductor, 
skin effect, proximity effect, and complexities of the power supply. 
All these factors contribute to the complexity of (pulsed) magnet
design, and typically make achieving the calculated performance
demanding in practice.

The use of pulsed magnets for vacuum-QED measurements at GW detectors was proposed
and evaluated in ref.~\cite{Doebrich2009}. The estimation of integration times
is more accurate in there, since the authors look at the full frequency
spectrum of a signal pulse whereas we made simplifying assumptions
in the estimation above.
However, the authors in~\cite{Doebrich2009} concentrated on the principal 
ideas of the approach, and did not consider a possibly realistic 
experimental setup.
One very optimistic assumption in their work is the estimation
of an aperture diameter (through which the laser beam passes) of the magnetic
field of order cm, which is the aperture available from pairs of pulsed
Helmholtz coils under development in the Dresden high-field 
laboratory~\cite{Doebrich2009}.
However, as shown above, a realistic assumption
is a necessary aperture diameter of order 10\,cm. 
This difference is the main single reason why the conclusion about the
feasibility of using pulsed magnets for QED measurements at GW detectors
is much more pessimistic in the estimation described in this section.

\subsection{Permanent magnets}
\label{permanent}

The development of permanent magnet materials has made significant progress
over the last decades, with the current maximum of a typical remanent magnetic flux
density around $B_r=1.3\,\rm T$ for neodymium-iron-boron ($Nd_2Fe_{14}B$) magnets.
With superposition arrangements of individual magnet domains it is possible
to obtain even larger magnetic field strengths, as for example with a 
Halbach array~\cite{Halbach1980}. If arranged in a \emph{Halbach cylinder}
configuration, a uniform magnetic field within a hollow cylinder can be 
obtained, with its field lines oriented perpendicular to the cylinder axis. 
The new design of the PVLAS experiment uses such Halbach cylinders
for the ellipsometric vacuum-QED measurement attempt~\cite{Valle2013}.

\begin{figure}[htbp]
\centering
\includegraphics[width=1.0\linewidth]{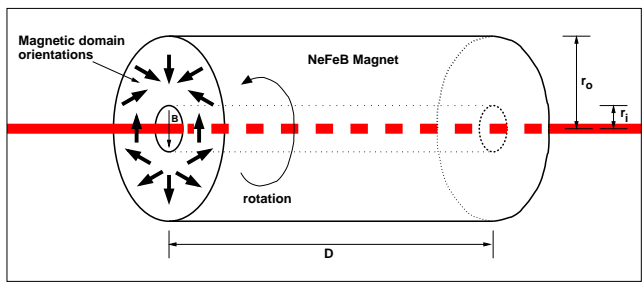}
\caption{Illustration of a Halbach cylinder with central opening, 
magnetized to yield a uniform magnetic field perpendicular to a laser
beam passing the cylinder.}
\label{HalbachCyl}
\end{figure}

Figure~\ref{HalbachCyl} shows a Halbach 
cylinder of length $D$ with an outer radius $r_o$ and a 
central opening of radius $r_i$. The magnetic domain orientations are depicted
on the left front face of the cylinder. The laser beam
passes through the central opening, and the cylinder would be rotated
around the laser beam axis, to provide spatial modulation of the magnetic field.

The magnetic field strength $B$ of a cylinder according to 
Figure~\ref{HalbachCyl} can be approximated by
\begin{equation}
\label{HalbachCylinder}
B = B_r \times ln \frac{r_o}{r_i}
\end{equation}
with $B_r$ being the remanent field strength of the magnet material.
Such a magnet can be used to spatially modulate the magnetic field by rotation
of the field around the laser beam, and thus measure $\Delta n_{\parallel - \perp}$.
A magnet with excitation $E=2.3\,\rm T^2 m$ could be constructed by choosing
$B_r=1.3\,\rm T$, $r_o/r_i = 2.2$, and $D=2.25\,\rm m$. Similar magnets have already
been fabricated for the new PVLAS experiment~\cite{Valle2013}, although with a
smaller inner radius $r_i$ than for this application, where we are aiming for 
$r_i = 55\,\rm mm$. 
Another similar magnet, for an aperture radius of 13.5\,mm,  
has been designed for the Q\&A experiment. This design is
similar to a Halbach cylinder, but also includes soft magnetic materials 
to increase the flux density towards the application region~\cite{Wang2004}.
However, given the total mass of the assembly, there seems to be 
no significant advantage over the standard Halbach cylinder design. 

An obvious large advantage of permanent- over electro-magnets is the fact that once
the magnet has been constructed, there is no shifting of energy in- and out of
the magnetic field required, and also no electrical power is dissipated in order to
generate the field. With the magnet described here, rotating at $f=25\,\rm Hz$
around a GW detector laser beam with a displacement noise of $\tilde n(\rm 50Hz) =
2 \times 10^{-20}\,\rm m/\sqrt{Hz}$, the integration time for n SNR=1 would again
be about 1 year.

\subsubsection*{Nested Halbach cylinders}

The disadvantage over electro-magnets is that in the setup 
using a single Halbach cylinder, only 
$\Delta n_{\parallel - \perp}$ can be measured. However, this
could be overcome by using two Halbach cylinders nested into each other. 
With such an arrangement it would be
possible to amplitude-modulate the magnetic field inside the inner cylinder,
simply by superposition of the fields of the two cylinders.
The orientation of the magnetic field lines stays constant
and thus $\Delta n_{\parallel}$ or $\Delta n_{\perp}$ can be measured individually.
The relative forces between the two cylinders are small in the ideal
case, where the outer fields are close to zero. Whether this approach would be
feasible in practice would need further investigation, and also depends
on the rotation speed of the magnets.

\section{A scenario to measure vacuum-QED effects with GW detectors}
\label{realistic}

\begin{figure*}[t]
\centering
\includegraphics[width=16cm]{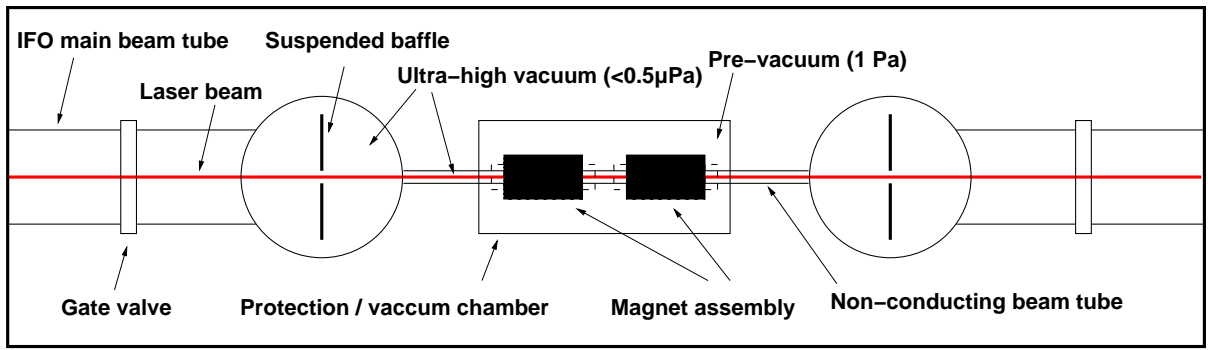}
\caption{Possible principle layout of a vacuum-QED measurement at the beam line
of a gravitational-wave detector. A section of the main beam tube is replaced
with a non-conducting section with small aperture. Seismically isolated
(suspended) baffles prevent light from being scattered at the interaction beam
line. Spatially or temporal modulated magnetic fields are generated by magnets
located around the interaction beam line. In case these magnets are solenoid
magnets to be rotated with high speed, additional pre-vacuum chambers might be
required, as depicted in the figure.}
\label{layout1}
\end{figure*}

Figure~\ref{layout1} shows a possible principal layout of a vacuum-QED
measurement at the beam line of a gravitational wave detector.
A section of the main beam tube is replaced
with a non-conducting section with small aperture. 
This tube should be electrically non-conducting to avoid attenuation of the 
usable magnetic field
by eddy-currents~\footnote{Simulations with FEMM show, that for an
amplitude modulated magnetic field at $f=25\,Hz$, a non-ferro-magnetic 
steel tube of 5\,mm wall strength would reduce the internal magnetic field
by about 50\,\% due to eddy currents.}. Eddy currents would also
heat the tube, and could lead to undesired mechanical forces.

Seismically isolated baffles are proposed
to prevent any light hitting the reduced beam tube, where scattering 
of laser light could
produce excess noise in the GW detector readout~\cite{Takahashi2004}.
The baffles should have a central opening that is slightly smaller
(e.g. by a few mm)
than that of the reduced beam tube, such that no light will hit
the beam tube in the interaction region with the magnetic field.
Obviously, the baffle opening diameter is then the limiting aperture
for the laser beam losses, and the amount by which this aperture
is smaller than the beam tube diameter determines how well the 
alignment of baffle and beam tube has to be set and maintained 
against each other.

The Cotton-Mouton effect (CME)~\cite{Cotton1905} of residual gas should be
sufficiently low if the total pressure is held at less than $0.5\,\rm \mu Pa$,
as far as the main constituents of air are concerned.
With estimated CMEs for molecular nitrogen and oxygen of 
$\Delta n_{\parallel - \perp}(N_2) \approx 2\times 10^{-12} / (10^5\,\rm Pa\,T^2)$ and
$\Delta n_{\parallel - \perp}(O_2) \approx 2\times 10^{-13} / (10^5\,\rm Pa\,T^2)$ as
taken from~\cite{Rizzo1997},
we approximate the CME for air ($78\,\rm \% N_2$ and $21\,\rm \% O_2$)
as $\Delta n_{\parallel - \perp}(air) \approx 6\times 10^{-13} / (10^5\,\rm Pa\,T^2)$.
With a total pressure of $0.5\,\rm \mu Pa$ we obtain a contribution
from residual air of $\Delta n_{\parallel - \perp}(air) \approx 3\times 10^{-24} / \rm T^2$,
just below the expected QED effect as given in equation~\ref{eqdeltanpp}.

The CME for water in the gas phase has been measured in \cite{Valle2014}
to $\Delta n_{\parallel - \perp}(H_2O) \approx 6.7\times 10^{-15} / (10^5\,\rm Pa\,T^2)$,
yielding a partial pressure of $\approx 60\,\rm \mu Pa$ to be
equal to the expected QED effect.

If a residual gas analyzer is used for permanent monitoring of the
partial pressures, and the CMEs of the residual gases are known sufficiently well, 
the estimated CME contributions can be subtracted from the vacuum QED signal
thus increasing the significance with which a vacuum QED effect 
can be isolated.

The main constructional challenge would be the assembly and precise alignment
of the reduced beam line setup and the suspended baffles.
Once this is done without degradation of the GW detector sensitivity,
the magnet experimental setup should not interfere with the GW detector
operations.
As discussed above, the best option seems to be permanent magnets
rotating around the beam axis.
As planned for the new PVLAS experiment, it is a good idea to use at least
two magnets. This opens the possibility to make null-measurements,
when the magnetic fields of the two magnets are kept perpendicular to each
other during rotation, thus testing for systematic errors due to false
signals. While more unlikely in a GW detector, such signals have been observed
in the ellipsometric experiments as described in section~\ref{introduction}.
In GW detectors the interaction region with the magnetic field would typically 
be at the middle of the beam tube, thus of order $\sim \rm km$ away from the
end stations holding the test-masses, which minimizes the risk of direct
interaction of the magnetic field with the test masses or other components
of the detection system.

Of course it is possible in principle that more systematic errors
would be discovered during the experiment. Systematic errors are commonly
the biggest unknown in high-precision experiments, and have been
slowing down the progress (not only) of other vacuum-QED projects.
For example, vibrations of the beam tube at twice the magnet rotation
frequency might be caused by (inhomogeneous) residual ferro-magnetic
contamination or diamagnetism / paramgnetism of the beam tube 
material~\footnote{I thank a reviewer for a comment to this effect.}.
These vibrations could couple to the laser beam (and thus may cause
a spurious signal) in principle, for example
if the shielding of the gaussian tail of the laser beam by the baffles
would not be sufficient.
However, it would be possible to measure the vibration of the beam tube, 
and for an additional null-test, one could excite a similar
beam tube motion as under magnet rotation, but without actually
rotating the magnets. Another possibility to exclude such an effect would be
to compare (QED-) measurements under different conditions of mechanical damping 
of the beam tube.

In order to estimate a limit for the vibration of the baffles at the 
signal frequency one has to make an assumption on the scatter of light
from the baffle into the main beam of the interferometer and compare
its contribution to a putative QED signal.
A very conservative 
estimate would be that a fraction $P_S = 10^{-10}$ of the main beam power
would be scattered into the fundamental Gaussian mode
\footnote{the exact amount strongly depends on the aperture size, baffle material,
and the micro-structure of the baffle edges.},
corresponding to a fraction $A_S = \sqrt{P_S} = 10^{-5}$ of the scattered
light field amplitude.
(Note that less than 1\,ppm of power should be lost due to forward 
scattering and clipping at the baffle, as discussed in section~\ref{magnets}.
Only a very small fraction of the clipped light can be scattered back into the
fundamental laser mode in principle.)
An excitation of $1\,\rm T^2m$ makes a signal of about $S \approx 10^{-24}\,\rm m$
(eq.\,\ref{eqdeltanpp} - \ref{pp-rms}) at frequency $f$, 
corresponding to a phase shift of 
$2 \pi / \lambda \times S \approx 6\times 10^{-18}\,\rm radian$ 
(with $\lambda=1064\,\rm nm$ being the wave-length of the light).
Therefore, the baffle motion at frequency $f$ 
should be less than $S/A_S \approx 10^{-19}\,\rm m$.
For a (hypothetical) motion of the baffle suspension
point of order $10^{-12}\,\rm m$ at frequency $f$,
one would thus need 7 orders of magnitude of isolation
from the baffle suspension.
If we take $f=50\,\rm Hz$ as signal frequency, this is achievable with
three stages of isolation (for example one passive stack/rubber pre-isolation
and a double pendulum suspension for the baffle.)
It is hard to predict what vibration level one may get at the suspension
point. However, this level can be measured precisely with accelerometers,
and the baffle suspension point could also
be artificially excited to estimate the amount of signal contribution
from scattering at the baffle.

\begin{figure}[htbp]
\centering
\includegraphics[width=1.0\linewidth]{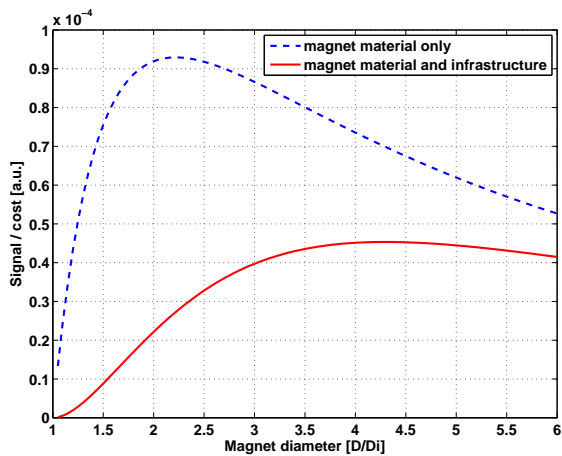}
\caption{Excitation per unit of cost as function of the radius
ratio $r_o/r_i$ for a Halbach cylinder as depicted in
Figure~\ref{HalbachCyl}. The maximum efficiency is around a ratio of 2.2
when only magnetic material and bearing cost are accounted for.
Additional cost of infrastructure that grows linear with length
of beam tube occupied by the setup can lead to an optimal ratio of 3.5-5.
}
\label{signalcost}
\end{figure}

Regarding the magnet design, a calculation of excitation per unit of material cost
shows that a ratio of $r_o/r_i = 2.2$ is close to the optimum,
as shown in Figure~\ref{signalcost}. 
For the bearing and rotational drive of the magnets, it seems best to use
friction-less magnetic bearings~\cite{Schweitzer2002}, 
which would be particularly helpful for the
long integration times needed. The cost of such bearings roughly scales with
the mass they can support, such that the cost optimization including bearings 
is the same as for magnet material only, as shown in Figure~\ref{signalcost}.
As an example, one could use two magnets with excitation
$E=1.2\,\rm T^2m$, and length $D=1.2\,\rm m$ each.
For a ratio $r_o/r_i=2.2$, and an inner radius of $r_i=55\,\rm mm$, the mass
of one magnet would be 328\,kg and it would cost around \$\,50000 at current 
material prices. For the two magnets, the integration time for a SNR=1 would be t=1.07\,years.
The situation gets better if the sensitivity of the GW detectors is
improved in the future. For example, a displacement noise of 
$8.5 \times 10^{-21}\,\rm m/\sqrt{Hz}$ at 50\,Hz might be reached by a potential upgrade of 
Advanced LIGO (see Figure~\ref{alldetectors}). Together with increasing the
number of magnets from 2 to 4, the integration time for SNR=1 would fall to t=17.7\,days,
such that after 3 years of operation a reasonably good SNR of 8 could be
achieved for the basic QED effect.

\section{Conclusion}
Laser-interferometric gravitational wave detectors currently under
construction or planned for the future offer the possibility of vacuum-QED
measurements. We have shown the principal feasibility of this approach
given the planned sensitivities and magnet technology,
and derived new estimates of measurement times. 
We have compared three different kinds of source field magnets and
conclude that from a realistic design perspective, permanent magnets
are the best, or even the only, option for the time being. The main implementation work
will come from the reduction of beam tube diameter, given the constraint to
not disturb the gravitational-wave measurement capability of the instrument.
Even if vacuum-QED measurements would be successful by ellipsometric
measurements within the next several years, the measurement of these effects with
GW detectors is still valuable since it is based on a different measured
quantity, the velocity shift of the light,
and also has the potential to measure parameters of exotic particle
models not accessible with ellipsometric measurements.

\section*{Acknowledgments}
The author thanks Harald L\"uck, Guido Zavattini, and Tobias Meier for useful
discussions, and Tobias Meier for an introduction to the FEMM program.
Thanks to Daniel Brown, Andreas Freise, and Vaishali Adya 
for help with simulations and thanks 
to Katherine Dooley and Benno Willke for valuable comments on the
manuscript.


\end{document}